# OscSNS: Precision Neutrino Measurements at the Spallation Neutron Source

**Heather Ray, for the OscSNS Collaboration**

University of Florida, Department of Physics, Gainesville, FL 32611

drhray@ufl.edu

**Abstract**. The Spallation Neutron Source (SNS), located at Oak Ridge Laboratory [1] in the United States, will be coming online over the next few years. In addition to producing fluxes of high-intensity neutrons, the interaction of the proton beam with the liquid mercury target produces copious pions. The $\pi^+$ and subsequent $\mu^+$ decay at rest, providing a neutrino beam comprising $\nu_\mu$, $\nu_e$, and anti-$\nu_\mu$ components. This neutrino beam is ideal for high-precision neutrino experiments. OscSNS [2] is a proposed multi-purpose experiment that will perform a search for light sterile neutrinos, search for beyond the Standard Model interactions using neutrino oscillations, and provide tests of Standard Model predictions through world-record precision neutrino cross section measurements. OscSNS plans to submit a full proposal for funding in 2009.

## 1. The Neutrino Beam

The SNS is the world's most intense pulsed accelerator-based neutron source. Spallation neutrons are created at the SNS through the interaction of a high-energy proton beam with a liquid mercury target. The proton beam is expected to reach 1.4 MW intensity at full power, at a 60 Hz rate. The SNS is designed to produce 695 ns width pulses; the SNS has already produced pulses less than 500 ns wide. This pulse width is significantly shorter than previous accelerator based neutrino experiments LSND [3] and MiniBooNE [4].

In addition to the spallation neutrons, the interaction of protons with the target also produces charged pions. These pions decay to produce a neutrino beam. Prior to decay, over 99.9% of the created $\pi^-$ are absorbed by the mercury target. The $\pi^+$ are brought to rest inside of the target, where their decay produces a mono-energetic 29.8 MeV $\nu_\mu$ and a $\mu^+$. The $\mu^+$ also decays at rest, providing a $\nu_e$ and anti-$\nu_\mu$ with known Michel energy distributions and end-point energy of 52.8 MeV.

The advantage of using a neutrino beam from mesons decaying at rest is that knowledge of the timing of the beam and the lifetime of the mesons allows for high suppression of cosmic ray backgrounds and isolation of a very high purity sample of the mono-energetic $\nu_\mu$. This type of beam also provides an extremely well defined neutrino flux. Two potential disadvantages to a decay at rest neutrino beam are that the beam is isotropic, and that the lower energy flux limits the choice of neutrino interactions you can examine. These two disadvantages are not concerns for OscSNS. The SNS produces an extremely high neutrino flux. With one year of run time OscSNS will far surpass the statistical size of several important neutrino measurement samples. Also, these low energy interactions are precisely the ones the OscSNS is interested in measuring.

## 2. OscSNS Detector

OscSNS is a 12 meter diameter spherical detector, lined with photomultiplier tubes (PMTs), and filled with mineral oil. The low energy neutrino interactions require the addition of a scintillator to the mineral oil. OscSNS will have two regions: an inner tank region, and an outer veto region. The inner region will be lined with 3262 PMTs, for a total of 25% coverage. OscSNS will be buried under 3 meters of dirt overburden, to help remove any residual cosmic ray events and beam induced neutrons as backgrounds.

The proposed location for OscSNS is 60 meters from the target, ~135 degrees in the backward direction with respect to the proton beam. This location will remove contributions to the neutrino beam coming from mesons that escape the target and decay in flight.

## 3. OscSNS Physics Program

The OscSNS experiment offers a complete physics program encompassing beyond the Standard Model searches as well as precision cross section measurements. OscSNS will produce neutrino oscillation searches using appearance as well as disappearance channels, search for light sterile neutrinos in a region of interest to astronomy and cosmology, perform world's best cross section measurements, and will provide an additional test of LSND and MiniBooNE low energy excess. Table 1 summarizes the physics program of OscSNS.

Table 1: Physics program of OscSNS.

| Cross Section Channels | Disappearance Channels ($\nu \to \nu_{sterile}$) | Appearance Channels |
|---|---|---|
| $\nu_e e^- \to \nu_e e^-$ | $\nu\,^{12}C \to\,^{12}C^*$ | $\nu_\mu \to \nu_e$: $\nu_e\,^{12}C \to e^{-\,12}N$ |
| $\nu_\mu e^- \to \nu_\mu e^-$, anti-$\nu_\mu e^- \to$ anti$\nu_\mu e^-$ | anti-$\nu\,^{12}C \to\,^{12}C^*$ | anti-$\nu_\mu \to$ anti-$\nu_e$: anti-$\nu_e\,^{12}C \to e^{+\,12}B$ |
| $\nu_\mu C \to \nu_\mu C$ | | anti-$\nu_\mu \to$ anti-$\nu_e$: anti-$\nu_e p \to e^+ n$ |
| $\nu_e\,^{12}C \to e^{-\,12}N$ | | |

### 3.1. Appearance Analyses

Appearance searches at the SNS will be performed using charged current quasi-elastic (CCQE) interactions. These searches will probe higher $\Delta m^2$ (0.01 to 1000 eV$^2$) and lower $\sin^2 2\theta$ (down to ~0.0001), in a region that will impact supernova and big bang nucelosynthesis modeling. The two appearance analyses employ a two-fold coincidence to separate candidate events from background sources. For the anti-$\nu_\mu \to$ anti-$\nu_e$ search the signal is a positron in coincidence with a 2.2 MeV $\gamma$: anti-$\nu_e p \to e^+ n$, followed by $np \to D\gamma$. This is an exceptionally clean channel. Intrinsic anti-$\nu_e$ backgrounds can only come from $\mu^-$, which are fairly non-existent at the SNS. For the $\nu_\mu \to \nu_e$ search the signal is an electron in coincidence with a positron, from the $\beta$ decay of the ground state of $^{12}$N: $\nu_e\,^{12}C \to e^{-\,12}N_{gs}$, followed by $^{12}N_{gs} \to\,^{12}C e^+ \nu_e$. The $\nu_\mu$ engaging in these oscillations are mono-energetic; this greatly limits the energy range for the produced positron and electron, and adds an additional layer of rejection of the intrinsic $\nu_e$ backgrounds.

### 3.2. Disappearance Analyses

Sterile neutrinos are the right handed partner to the Standard Model neutrinos. Sterile neutrinos don't engage in the Weak interaction. They can only be observed via their mixing with their left handed Standard Model partners. OscSNS will use the super-allowed neutral current (NC) interaction to search for sterile neutrinos, $\nu\,^{12}C \to\,^{12}C^*$, $^{12}C^* \to\,^{12}C\gamma$. The Carbon emits a very distinctive 15.11 MeV gamma that will be used to identify these events.

Theoretical arguments claim a light sterile neutrino should have a high mixing angle with Standard Model neutrinos [5], but offer a wide range of possible angles. For sensitivity estimates we want to use an oscillation probability that represents the average of possible values, covering a wide range of masses and mixing angles. We also need to consider that the energy spectra of incident neutrinos will produce an L/E value varying between 6 and 1. The probability of 0.26% represents such an average, and is used for both disappearance and appearance estimates and sensitivity curves.

3.3. Cross Section Analyses

The flagship cross section analyses of OscSNS are the elastic scattering $\nu_e e^- \to \nu_e e^-$ (NC and CC), NC $\nu_\mu e^- \to \nu_\mu e^-$, NC anti-$\nu_\mu e^- \to$ anti$\nu_\mu e^-$, NC $\nu_\mu C \to \nu_\mu C$, and the CC $\nu_e{}^{12}C \to e^{-12}N$ interactions.

The current world's best measurement of the $\nu_e e^- \to \nu_e e^-$ interaction arises in a sample of only 191 events [6], and has 17% total error. OscSNS will far surpass this measurement in statistics and total uncertainty, with only one year of run time. The Super-Kamiokande [7] experiment collected over 100,000 of these events. However, they did not know their flux. They had to assume a cross section for the $\nu_e e^- \to \nu_e e^-$ events to quantify their observed neutrino oscillations. OscSNS will be able to perform the world's best direct measurement of this interaction cross section.

The NC $\nu_\mu C \to \nu_\mu C$ interaction has been measured by the KARMEN experiment to be $3.2 \pm 0.5 \pm 0.4 * 10^{-42}$ cm$^2$ [8]. This measurement was performed using only 86 $\nu_\mu$ events. While this result is consistent with theory [9], it has a 20% total error, half of which is due to statistics. OscSNS will collect 1,342 events in only one year of run time, and is expected to have smaller systematic errors. This will allow for the world's most precise cross section measurement; any deviations from theory could indicate the presence of sterile neutrinos or other new physics.

For the CC $\nu_e{}^{12}C \to e^{-12}N$ measurement, we are able to use the entire spectrum of intrinsic $\nu_e$, in contrast to the oscillation search, where we only consider the neutrino flux in the first 500 ns of the beam spill.

3.4. Event Rate and Expected Sensitivity

The following event rate and sensitivity estimates consider neutrino backgrounds only. The cosmic ray background is assumed to be negligible in these studies. Event rate estimates take into account the reduced volume of the detector (here expected to be 5 meter fiducial radius), and a 50% reduction due to detector efficiency. Table 2 presents the expected rates for the various analyses, for 1 year of run time. Figures 1 and 2 present the expected sensitivity for the two appearance searches, for 3 years of run time.

**4. Summary**

The SNS complex is in the final stages of commissioning. The mechanism that produces spallation neutrons also produces a copious amount of neutrinos. The beam structure of the SNS allows for excellent and simultaneous measurements in neutrino and anti-neutrino modes. The neutrino flux from the SNS has a well known energy spectrum, necessary for precision measurements. The OscSNS experiment may be built for around 10M dollars, and provides a multi-faceted physics program. OscSNS will achieve the world's best neutrino oscillation sensitivities for $\nu_\mu \to \nu_e$ oscillations, anti-$\nu_\mu \to$ anti-$\nu_e$ oscillations, and $\nu_\mu \to \nu_{sterile}$ oscillations at a $\Delta m^2$ scale of greater than 0.1 eV$^2$. OscSNS will also perform the world's best measurement of $\nu_e e^- \to \nu_e e$ and $\nu_e C \to e^- N$ scattering. In addition, the SNS has a future upgrade planned that entails adding a second target station, creating the interesting situation of two baselines for OscSNS.

Table 2: Expected event rates for a 5 meter radius fiducial volume detector, located at the SNS 60 meters from the target, at ~135 degrees in the backward direction from the proton beam. All event rates account for a 50% detector efficiency, and are in units of expected events per year. Appearance signal estimates assume a 0.26% oscillation probability.

| Cross Section Channels | | Disappearance Channels | | Appearance Channels | | |
|---|---|---|---|---|---|---|
| | Event Rate | | Event Rate | | Background | Signal |
| $\nu_e e^- \to \nu_e e^-$ | 1,302 | $\nu\,^{12}C \to ^{12}C^*$ anti-$\nu\,^{12}C \to ^{12}C^*$ | 6,322 | $\nu_e\,^{12}C \to e^-\,^{12}N$ | 16 | 6 |
| $\nu_\mu e^- \to \nu_\mu e^-$ | 204 | $\nu_\mu\,^{12}C \to ^{12}C^*$ | 1,342 | anti-$\nu_e\,^{12}C \to e^+\,^{12}B$, anti-$\nu_e p \to e^+ n$ | 79 | 253 |
| anti-$\nu_\mu e^- \to$ anti$\nu_\mu e^-$ | 217 | | | | | |
| $\nu_\mu C \to \nu_\mu C$ | 1,342 | | | | | |
| $\nu_e\,^{12}C \to e^-\,^{12}N$ | 4,616 | | | | | |

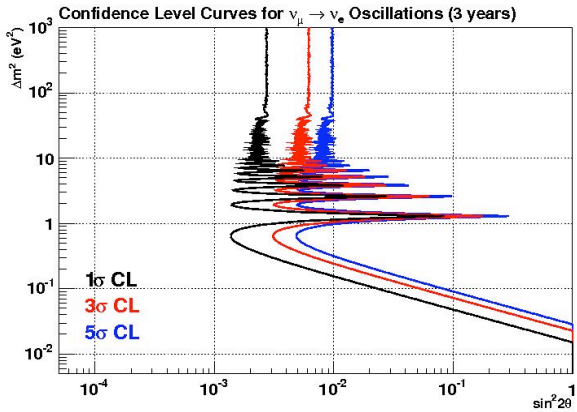

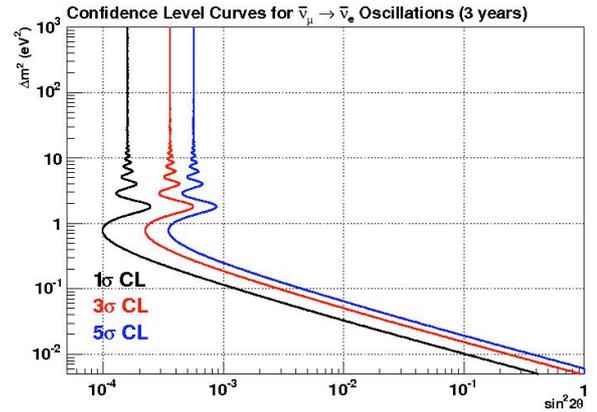

**Figure 1.** OscSNS sensitivity for $\nu_\mu \to \nu_e$ oscillations, for three years of run time. The 1σ confidence level band is furthest to the left; 5σ is furthest to the right.

**Figure 2.** OscSNS sensitivity for anti-$\nu_\mu \to$ anti-$\nu_e$ oscillations, for three years of run time. The 1σ confidence level band is furthest to the left; 5σ is furthest to the right.